\documentclass{article}

\usepackage[nonatbib,final]{neurips_2023_ml4ps}
\usepackage{amsmath}
\usepackage{graphicx}
\usepackage{multirow}
\usepackage{tabularx}
\usepackage{booktabs}
\usepackage[utf8]{inputenc} 

\usepackage[%
    style=phys,
    biblabel=brackets,
    eprint=true,
    pageranges=false,
]{biblatex}
\AtEveryBibitem{\clearfield{month}}
\usepackage[T1]{fontenc}    
\usepackage{hyperref}       
\usepackage{url}            
\usepackage{booktabs}       
\usepackage{amsfonts}       
\usepackage{nicefrac}       
\usepackage{microtype}      
\usepackage{xcolor}         
\usepackage{xspace}

\renewcommand{\vec}[1]{\mathbf{#1}}

\newcommand{\pt}{\ensuremath{p_{\mathrm{T}}}\xspace}

\newcommand{\etarel}{\ensuremath{\eta^\mathrm{rel}}\xspace}
\newcommand{\phirel}{\ensuremath{\phi^\mathrm{rel}}\xspace}
\newcommand{\ptrel}{\ensuremath{\pt^\mathrm{rel}}\xspace}

\newcommand{\wassm}{\ensuremath{W_1^\mathrm M}\xspace}
\newcommand{\wassppt}{\ensuremath{W^{\ptrel}_{1p}}\xspace}

\title{Induced Generative Adversarial Particle Transformers}

\author{%
  Anni Li\thanks{Equal contribution.} \quad Venkat Krishnamohan$^{\ast}$ \quad Raghav Kansal\thanks{Also at Fermi National Accelerator Laboratory.}\\
  \textbf{Rounak Sen \quad Steven Tsan \quad Zhaoyu Zhang \quad Javier Duarte} \\
  University of California, San Diego\\
  La Jolla, CA 92093, USA\\
  \texttt{\{a5li,vkrishnamohan,rkansal,r2sen,stsan,zhz057,jduarte\}@ucsd.edu}
}

\begin{document}

\begin{flushright}
FERMILAB-CONF-23-751-CMS-PPD
\end{flushright}
\maketitle

\begin{abstract}
  In high energy physics (HEP), machine learning methods have emerged as an effective way to accurately simulate particle collisions at the Large Hadron Collider (LHC). 
  The message-passing generative adversarial network (MPGAN) was the first model to simulate collisions as point, or ``particle'', clouds, with state-of-the-art results, but suffered from quadratic time complexity.
  Recently, generative adversarial particle transformers (GAPTs) were introduced to address this drawback; however, results did not surpass MPGAN.
  We introduce induced GAPT (iGAPT) which, by integrating ``induced particle-attention blocks'' and conditioning on global jet attributes, not only offers linear time complexity but is also able to capture intricate jet substructure, surpassing MPGAN in many metrics.
  Our experiments demonstrate the potential of iGAPT to simulate complex HEP data accurately and efficiently.
\end{abstract}

\section{Introduction}

In high energy physics (HEP), jets---sprays of particles produced at high energy collisions at the Large Hadron Collider (LHC), serve as a vital bridge between theoretical research and experimental observations. 
The intricate patterns in jets make them invaluable for understanding quantum chromodynamics (QCD), identifying rare particles like the Higgs boson, and conducting quantitative analyses of particle properties and interactions. 
Machine learning methods, such as generative adversarial networks (GANs), have recently emerged as effective and efficient approaches to simulating such high-energy collisions, reducing the computational burden of traditional methods.
The message-passing GAN (MPGAN)~\cite{kansal2021particle}, in particular, represented a significant leap forward, being able to capture the complex global structure of jets and handle variable-sized particle clouds using fully-connected graph neural networks.
However, this fully-connected nature means its memory and time complexity scale quadratically with the number of particles per jet.

The generative adversarial particle transformer (GAPT)~\cite{kansal2023evaluating} was recently introduced to improve the efficiency of MPGAN using self-attention, but GAPT does not match MPGAN's performance.
Others have also studied normalizing flows~\cite{Orzari:2023zrh}, other types of transformers~\cite{Kach:2022uzq,Kach:2023rqw}, diffusion models~\cite{Leigh:2023toe,Mikuni:2023dvk}, and equivariant networks~\cite{Buhmann:2023pmh} for the same generative task.
In this paper, we introduce the induced GAPT (iGAPT), featuring ``induced particle attention blocks'' incorporating physics-informed inductive biases of jets, to offer linear time complexity and improve output fidelity.
We investigate its performance and timing compared to MPGAN and alternative architecture choices on high-momentum jets.

\section{Architecture}
\label{sec:arch}

\begin{figure}[ht]
    \centering
    \includegraphics[width=1\linewidth]{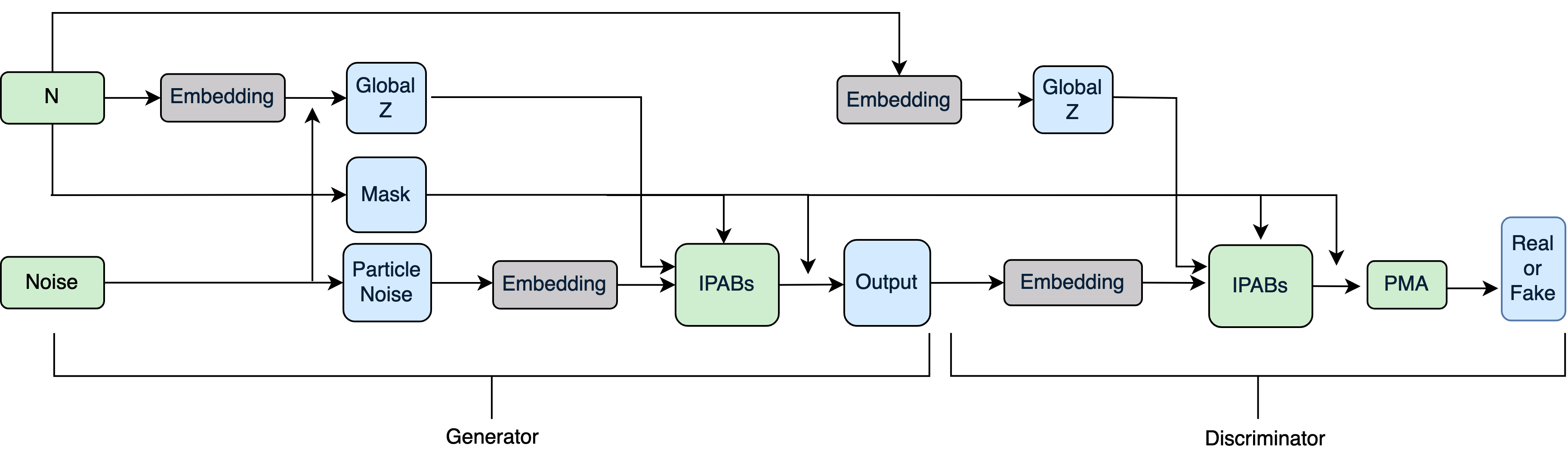}
    \caption{Illustration of the generation and discrimination process of iGAPT.}
    \label{mainstructure}
\end{figure}

The iGAPT model, illustrated in Fig.~\ref{mainstructure}, is a GAN with the generator and discriminator comprising ``induced particle attention blocks'' (IPABs).
On top of permutation invariance and point-cloud representations, as in MPGAN and GAPT, the key inductive bias we experiment with in iGAPT is maintaining a global vector through the generation and discrimination processes, $\vec{z}$, which implicitly represents global jet features.
IPABs and different ways of incorporating $\vec{z}$ into the attention process are described below.

The generation process starts with sampling random Gaussian noise and a particle multiplicity $N$ from the true distribution, which is transformed via a learned linear embedding layer.
The noise has two components: a set of $N_\mathrm{max}$ vectors representing initial particle features, and a single vector representing initial jet features.
$N_\mathrm{max}$ is the maximum number of particles per jet we want to simulate, and the number of initial particle and jet features is a hyperparameter we tune.
The jet noise is added to the embedded $N$ to produce $\vec{z}$, which is then transformed along with the particle noise via multiple IPABs to output a generated jet.

The discrimination process starts with a generated or real jet, and the sampled or true jet multiplicity $N$.
This is again transformed via a learned embedding layer to produce the 
$\vec{z}$ conditioning vector for the discriminator, which along with the input jet are processed through IPABs producing an intermediate jet representation.
The constituents of this jet are aggregated in a permutation-invariant manner using a pooling by multihead attention (PMA) layer, as introduced in~\cite{lee2019set}, the output of which is finally fed into a linear layer to classify the jet as real or fake.

Attention blocks require fixed multiplicity, however, jets naturally have a variable number of particle constituents.
To handle this, we zero-pad all jets to $N_\mathrm{max}$ particles and use masked attention in every block to ignore these.

\paragraph{Attention Blocks}
\label{sec:attentionblocks}
\begin{figure}[ht]
    \centering
    \includegraphics[width=0.5\linewidth]{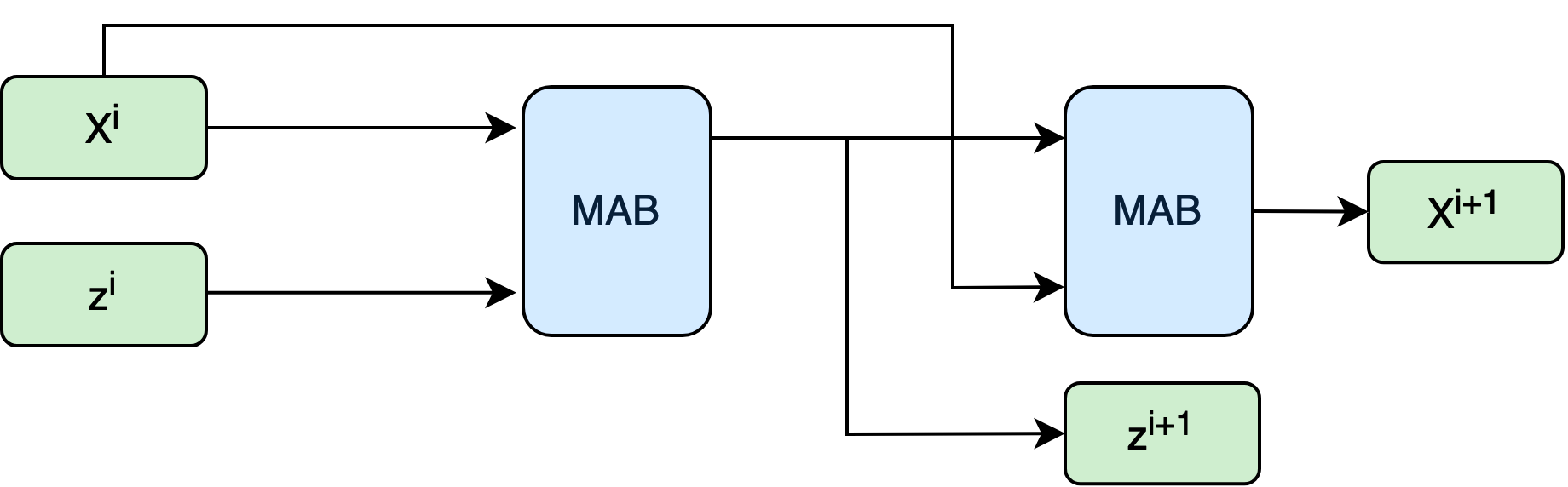}
    \caption{Illustration of an induced particle attention block (IPAB).}
    \label{IPABfigure}
\end{figure}

The original GAPT model uses self-attention blocks (SABs) introduced for set transformers~\cite{lee2019set}, a form of multi-head attention blocks (MABs) with the input attending itself --- i.e., each particle attends to every other in the jet to update its features.
This has the benefit of global, fully connected interactions between particles; however, it also means quadratic scaling with the particle multiplicity, which is undesirable.

To counter this, Ref.~\cite{lee2019set} also proposed ``induced'' SABs (ISABs), where the input set is first attended to by $M$ learnt inducing vectors via a MAB, outputting an intermediate, compressed representation.
This representation is then attended to by the original set to produce the new output.
This retains the global interactions between particles while achieving $\mathcal O(NM)$ scaling---linear in the particle multiplicity.
ISABs have been used in generative adversarial set transformers (GAST)~\cite{stelzner2020generative}, yielding high quality results on computer vision datasets.
GAST incorporates similar global set features $\vec{z}$ by concatenating them to each particle before every ISAB operation.

In iGAPT, we experiment with an alternative version of conditioning, which we call ``induced particle attention''.
Here, $\vec{z}$ is directly used as the inducing vector in an ISAB, and is continuously updated through the intermediate, induced attention outputs.
Explicitly, as illustrated in Fig.~\ref{IPABfigure}, the $i$th IPAB receives as input the jet $\vec{x}^{i}$ and global features $\vec{z}^{i}$ from the previous block, after which $\vec{x}^{i}$ is first attended to by $\vec{z}^{i}$ to output updated features $\vec{z}^{i+1}$, and then conversely $\vec{z}^{i+1}$ is attended to by $\vec{x}^{i}$ to output the updated jet $\vec{x}^{i+1}$.
This is interpreted as a way to update and learn the global jet features, such as mass and \pt, along with the individual particle features, and allow both to interact in each attention layer.
An additional and significant advantage is that this induced attention operation involves only one inducing vector --- $\vec{z}$, hence $M = 1$ and we achieve $\mathcal O(N)$ computational complexity.

\section{Experiments}

\paragraph{Dataset}

We perform tests on the JetNet dataset~\cite{kansal_raghav_2022_6975118} using the associated library~\cite{Kansal_JetNet_2023,kansal2021particle}.
The JetNet dataset contains simulated high transverse-momentum (\pt) jets, originating from gluons, light quarks, top quarks, incorporating detector effects emulating those of the CMS detector at the LHC.
We test the iGAPT model on these different classes, simulating only the 30 as well as the 150 highest \pt particles per jet.
This complex task involves simulating both the initial parton showering and hadronization from an individual quark or gluon, as well as the proceeding detector response.
Each jet is represented as a set, or point cloud, of its constituent particles, which are described by three features: their angular coordinates (\etarel, \phirel), and \pt, relative to the jet.

\paragraph{Training and Hyperparameters}


The iGAPT and GAST models were trained for a maximum of 6000 epochs on a single NVIDIA RTX1080 GPU using the RMSProp optimizer.
The training time and batch size for 30- and 150-particle gluon jets is shown in Table~\ref{tab:times_gluon}.
We use a two-time update rule~\cite{TTUR}, with learning rates of $0.5 \times 10^{-4}$ and $1.5 \times 10^{-4}$ for the generator and discriminator, respectively.
The discriminator is regularized with dropout with a probability of 0.5 and layer normalization.
We use a LeakyReLU activation after every linear layer, except the final generator (tanh) and discriminator (sigmoid) outputs.
After hyperparameter tuning, we settle on on 3 and 6 ISAB layers for the generator and discriminator respectively, and 4 and 8 IPAB layers for the iGAPT model.
We use 16 and 128 initial particle and jet features, respectively, in iGAPT.
GAST uses 20 inducing vectors for its ISABs.

\paragraph{Evaluation}

We evaluate the performance of each model both visually, comparing feature distributions between the real and generated samples, and using quantitative evaluation metrics.
Reference~\cite{kansal2023evaluating} found the Fr\'{e}chet physics distance (FPD), kernel physics distance (KPD), and 1-Wasserstein distances ($W_1$) between individual particle and jet features to be sensitive metrics to comprehensively evaluate the physics performance of generative models.
FPD and KPD, in particular, consider a set of 36 energy flow polynomials (EFPs)~\cite{komiske_efps}, which collectively form a basis for all infrared- and collinear-safe observables, and hence capture significant information about the jet substructure.
We report the FPD, KPD, and $W_1$ scores for the particle \pt (\wassppt) and jet mass (\wassm) and their errors following the procedures outlined in Ref.~\cite{kansal2023evaluating} using the JetNet library for each generative model, as well as between independent samples of real jets to measure a baseline, ``truth'' score.

\paragraph{Results}

\begin{figure}
    \centering
    \includegraphics[width=\textwidth]{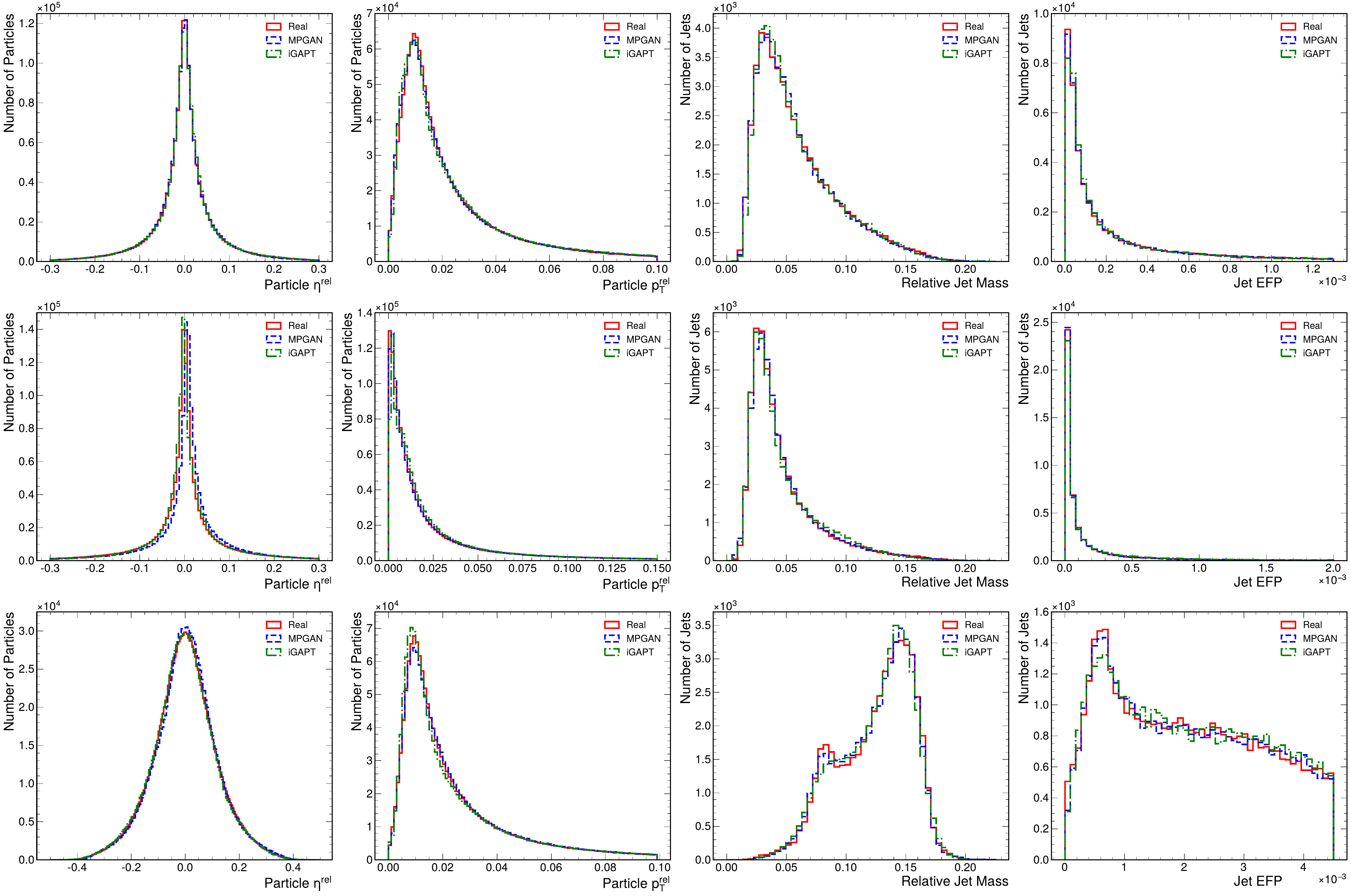}
    \caption{Low-level particle feature distributions (far left and center left) and high-level jet feature distributions (center right and far right) for the real data (red), MPGAN-generated data (blue), and iGAPT-generated data (green), for 30-particle gluon (top row), light quark (middle), and top quark jets (bottom).
    A sample $d = 4$ energy flow polynomial~\cite{komiske_efps} is plotted in the rightmost column.
    }
    \label{fig:feature_distributions_30}
\end{figure}

We test the iGAPT model on 30-particle gluon, light quark, and top quark jets, and on 150-particle gluon jets.
We compare this with the MPGAN and GAPT models provided in their respective publications, as well as as our own trainings of GAST models.
Out of all our trainings, we select the iGAPT and GAST models with the lowest FPD score, as advocated in Ref.~\cite{kansal2023evaluating} due to its high sensitivity to common types of mismodelling.
Comparisons of real and generated feature distributions are shown in Fig.~\ref{fig:feature_distributions_30} and Appendix~\ref{app:features} for 30 and 150 particles, respectively, demonstrating high fidelity results by iGAPT.
The truth and model scores for the aforementioned evaluation metrics are listed in Table~\ref{tab:finaleval}.
We note that MPGAN models for 150-particle jets are not provided because of their poor time complexity. 
We observe first that iGAPT largely outperforms the GAPT and GAST models, validating our physics-informed approach to its architecture.
It also outperforms MPGAN on 30-particle jets on many jet types and metrics, notably on FPD for all jet types.
Important we recognize that no model is perfectly consistent with the true, baseline distributions, indicating room for improvement.

\begin{table}
    \centering
    \caption{Evaluation metrics for different jet types and models.
    The best-performing model on each metric and jet type is highlighted in bold.}
    \label{tab:finaleval}
    \begin{tabular}{l|lcccc}
        \toprule
        & Model & \wassppt (\(10^{-3}\)) & \wassm (\(10^{-3}\)) & FPD (\( 10^{-3}\)) & KPD (\(10^{-6}\)) \\
        \midrule
\multirow{5}{*}{Gluon (30)}
 & Truth & $0.14 \pm 0.06$ & $0.46 \pm 0.08$ & $0.14 \pm 0.04$ & $1.8 \pm 11.9$\\ 
 & MPGAN & $0.27 \pm 0.02$ & $0.7 \pm 0.3$ & $0.41 \pm 0.09$ & $\mathbf{0 \pm 8}$\\ 
 & GAPT & $\mathbf{0.25 \pm 0.07}$ & $1.0 \pm 0.2$ & $0.46 \pm 0.06$ & $5 \pm 3$\\ 
 & GAST & $0.8 \pm 0.1$ & $0.7 \pm 0.2$ & $0.40 \pm 0.05$ & $7.0 \pm 10.3$\\ 
 & iGAPT & $0.76 \pm 0.07$ & $\mathbf{0.7 \pm 0.1}$ & $\mathbf{0.29 \pm 0.04}$ & $3 \pm 5$\\ \cline{1-6}
\multirow{5}{*}{Light quark (30)}
 & Truth & $0.21 \pm 0.05$ & $0.5 \pm 0.2$ & $0.09 \pm 0.03$ & $-3 \pm 3$\\ 
 & MPGAN & $\mathbf{0.41 \pm 0.07}$ & $\mathbf{0.5 \pm 0.1}$ & $1.9 \pm 0.2$ & $\mathbf{1.7 \pm 15.1}$\\ 
 & GAPT & $2.74 \pm 0.09$ & $2.54 \pm 0.05$ & $4.03 \pm 0.06$ & $96 \pm 9$\\ 
 & GAST & $1.2 \pm 0.1$ & $1.8 \pm 0.2$ & $0.65 \pm 0.07$ & $27.0 \pm 11.7$\\ 
 & iGAPT & $1.89 \pm 0.04$ & $1.2 \pm 0.3$ & $\mathbf{0.51 \pm 0.07}$ & $12 \pm 7$\\  \cline{1-6}
\multirow{5}{*}{Top quark (30)}
 & Truth & $0.20 \pm 0.05$ & $0.7 \pm 0.2$ & $0.07 \pm 0.03$ & $-16 \pm 2$\\ 
 & MPGAN & $0.44 \pm 0.08$ & $\mathbf{0.5 \pm 0.1}$ & $2.8 \pm 0.2$ & $14.7 \pm 12.9$\\ 
 & GAPT & $\mathbf{0.34 \pm 0.02}$ & $1.9 \pm 0.2$ & $0.43 \pm 0.03$ & $25.4 \pm 28.8$\\ 
 & GAST & $1.16 \pm 0.08$ & $1.5 \pm 0.2$ & $0.30 \pm 0.05$ & $\mathbf{-2.4 \pm 17.2}$\\ 
 & iGAPT & $0.54 \pm 0.04$ & $0.9 \pm 0.3$ & $\mathbf{0.25 \pm 0.03}$ & $-0.6 \pm 14.1$\\  \cline{1-6}
\multirow{4}{*}{Gluon (150)}
 & Truth & $0.09 \pm 0.03$ & $0.7 \pm 0.2$ & $0.10 \pm 0.03$ & $0.5 \pm 10.5$\\ 
 & GAPT & $0.77 \pm 0.03$ & $\mathbf{1.1 \pm 0.3}$ & $22.0 \pm 0.1$ & $62.5 \pm 11.1$\\ 
 & GAST & $0.68 \pm 0.05$ & $3.7 \pm 0.3$ & $3.60 \pm 0.06$ & $\mathbf{47.7 \pm 13.8}$\\ 
 & iGAPT & $\mathbf{0.66 \pm 0.03}$ & $4.4 \pm 0.7$ & $\mathbf{2.99 \pm 0.06}$ & $158.1 \pm 37.9$\\ 
        \bottomrule
    \end{tabular}
\end{table}

\paragraph{Timing} 

A key benefit of iGAPT is its improved time complexity over MPGAN.
This is demonstrated in Table~\ref{tab:times_gluon}, which shows the training and generation times for each model for 30 particle jets using the largest batch size possible on an NVIDIA 1080 GPU, with iGAPT outperforming MPGAN by a factor of 3.5.
MPGAN is computationally challenging to extend to 150 particles, hence timing information is not provided; in contrast, iGAPT's training and generation times scale well with the number of particles.
Finally, we note that the ``true'' generation time per jet is approximately 50\,ms~\cite{kansal2021particle}, thus iGAPT represents more than a factor of 100 speed up.


\begin{table}
\centering
\caption{Timing measurements for MPGAN and iGAPT, measured on an NVIDIA 1080 GPU.}
\label{tab:times_gluon}
\begin{tabular}{l|lccccc}
\toprule
Jet type & Model & Training time & & Generation time & & Batch size \\
 & & (s/epoch) & & (µs / jet) & \\
\midrule
\multirow{2}{*}{Gluon, \(n=30\)} 
& MPGAN & $193$ & & $142$ & & 512 \\
& iGAPT & \textbf{$31$} & & \textbf{$40$} & & 4096 \\
\midrule
\multirow{1}{*}{Gluon, \(n=150\)} 
& iGAPT & $267$ & & $315$ & & 512 \\
\bottomrule
\end{tabular}
\end{table}

\section{Conclusion}
We presented the induced generative adversarial particle transformer (iGAPT), which utilizes a new induced particle attention mechanism to achieve high-fidelity simulations of jets in high energy physics. 
By incorporating physics-informed inductive biases and conditioning on global jet features, iGAPT demonstrates not only competitive performance with previous state-of-the-art models on 30-particle gluon, light quark, and top quark jets, but also significantly better computational efficiency and time complexity. 
Furthermore, iGAPT exhibits strong potential to simulate larger numbers of particles, establishing its viability in applications to realistic event collisions at the LHC.
However, we note room for improvement in iGAPT's performance with respect to the baseline scores, and future work may explore further optimizations of its training and performance.


\begin{ack}
This work was supported by the U.S. Department of Energy (DOE) Award Nos. DE-SC0021187 and DE-SC0021396 (FAIR4HEP) and the U.S. National Science Foundation (NSF) Cooperative Agreement OAC-2117997 (A3D3).
A.~L. was partially supported by an IRIS-HEP fellowship through the U.S. National Science Foundation (NSF) under Cooperative Agreement OAC-1836650.
R.~K. was partially supported by the LHC Physics Center at Fermi National Accelerator Laboratory, managed and operated by Fermi Research Alliance,
LLC under Contract No. DE-AC02-07CH11359 with the DOE.
\end{ack}


\printbibliography

\appendix

\section{Feature distributions of 150-particle jets}
\label{app:features}

\begin{figure}[ht]
    \centering
    \includegraphics[width=\textwidth]{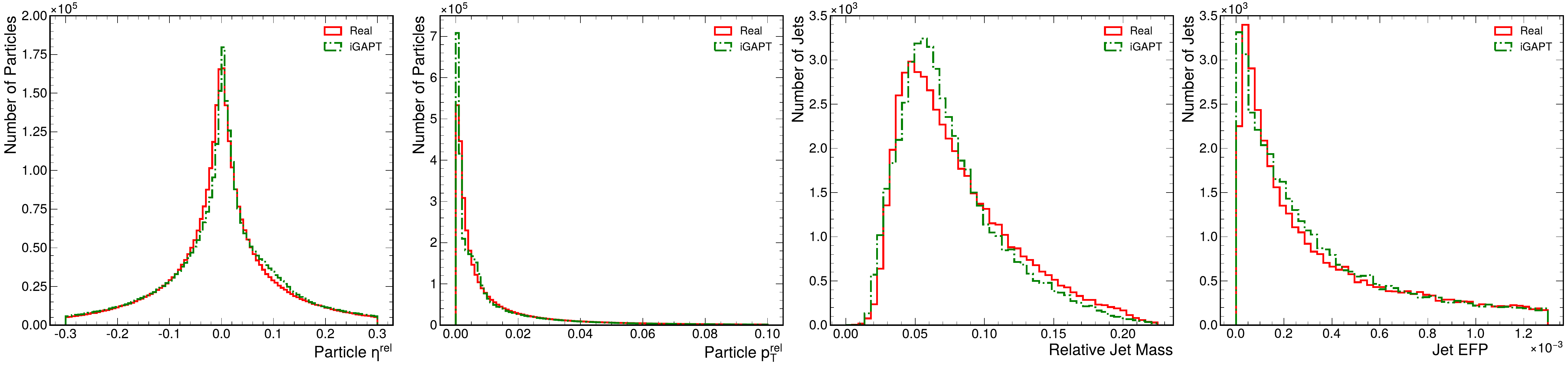}
    \caption{Low-level particle feature distributions (far left and center left) and high-level jet feature distributions (center right and far right) for the real data (red), MPGAN-generated data (blue), and iGAPT-generated data (green), for 150-particle gluon jets.
    A sample $d = 4$ energy flow polynomial~\cite{komiske_efps} is chosen in the rightmost column.
    }
    \label{fig:feature_distributions_150}
\end{figure}

The distributions of real and generated particle and jet features for 150-particle gluon jets by MPGAN and iGAPT are shown in Fig.~\ref{fig:feature_distributions_150}.

\end{document}